\documentclass[epj,a4,twocolumn]{webofc}
\pdfoutput=1
\usepackage[varg]{txfonts}   
\woctitle{CGS15}
\begin{document}
\title{Sensitivity Increases for the TITAN Decay Spectroscopy Program}
%
%

\author{K.G.~Leach\inst{1,2}\fnsep\thanks{\email{kleach@triumf.ca}} \and
        A.~Lennarz\inst{1,3}\and
        A.~Grossheim\inst{1}\and
        C.~Andreoiu\inst{2}\and
        J.~Dilling\inst{1,4}\and
        D.~Frekers\inst{3}\and
        M.~Good\inst{1}\and
        S.~Seeraji\inst{2}
}

\institute{TRIUMF, 4004 Wesbrook Mall, Vancouver BC, V6T 2A3, Canada
\and
Department of Chemistry, Simon Fraser University, Burnaby BC, V5A 1S6, Canada
\and
Instit\"ut f\"ur Kernphysik, Westfalische-Wilhelms-Universit\"at M\"unster, D-48149, Germany
\and
Department of Physics and Astronomy, University of British Columbia, Vancouver, BC, V6T 1Z1, Canada
          }



\abstract{%
The TITAN facility at TRIUMF has recently initiated a program of performing decay spectroscopy measurements in an electron-beam ion-trap (EBIT).  The unique environment of the EBIT provides backing-free storage of the radioactive ions, while guiding charged decay particles from the trap centre via the strong magnetic field.  This measurement technique is able to provide a significant increase in detection sensitivity for photons which result from radioactive decay.  A brief overview of this device is presented, along with methods of improving the signal-to-background ratio for photon detection by reducing Compton scattered events, and eliminating vibrational noise.
}
\maketitle
\section{Introduction}
TRIUMF's Ion Trap for Atomic and Nuclear Science (TITAN)~\cite{Dil06} is located at the Isotope Separator and Accelerator (ISAC) facility at TRIUMF, in Vancouver, BC, Canada.  TRIUMF-ISAC produces rare-isotope beams (RIBs) using the isotope separation on-line (ISOL) technique~\cite{Blu13}, by generating spallation reactions on thick production targets using an up to 100 $\mu$A beam of 500 MeV protons.  The mass-selected, continuous beam of radioactive ions is delivered at low energies ($\sim20$~keV) to a suite of experimental facilities for both cooled- and stopped-beam experiments~\cite{Dil14}.

The TITAN system consists of three ion traps: a buffer-gas-filled radio-frequency-quadrupole (RFQ)~\cite{Bru12}, a 3.7~T Penning-trap for precision mass measurements (MPET)~\cite{Bro12a}, and an electron-beam ion trap (EBIT)~\cite{Lap10} which provides highly charged ions (HCIs) for the Penning trap.  The addition of two new components to the TITAN system is planned for the near future; a cooler Penning-trap (CPET) to sympathetically cool HCIs with electrons or protons~\cite{Sch13}, and a multi-reflection time-of-flight (MR-ToF) isobar separator~\cite{Pla13}.  Both of these devices have been delivered to TRIUMF and should be installed in the TITAN system within the next year.
\begin{figure}[t!]
\centering
\includegraphics[width=\columnwidth]{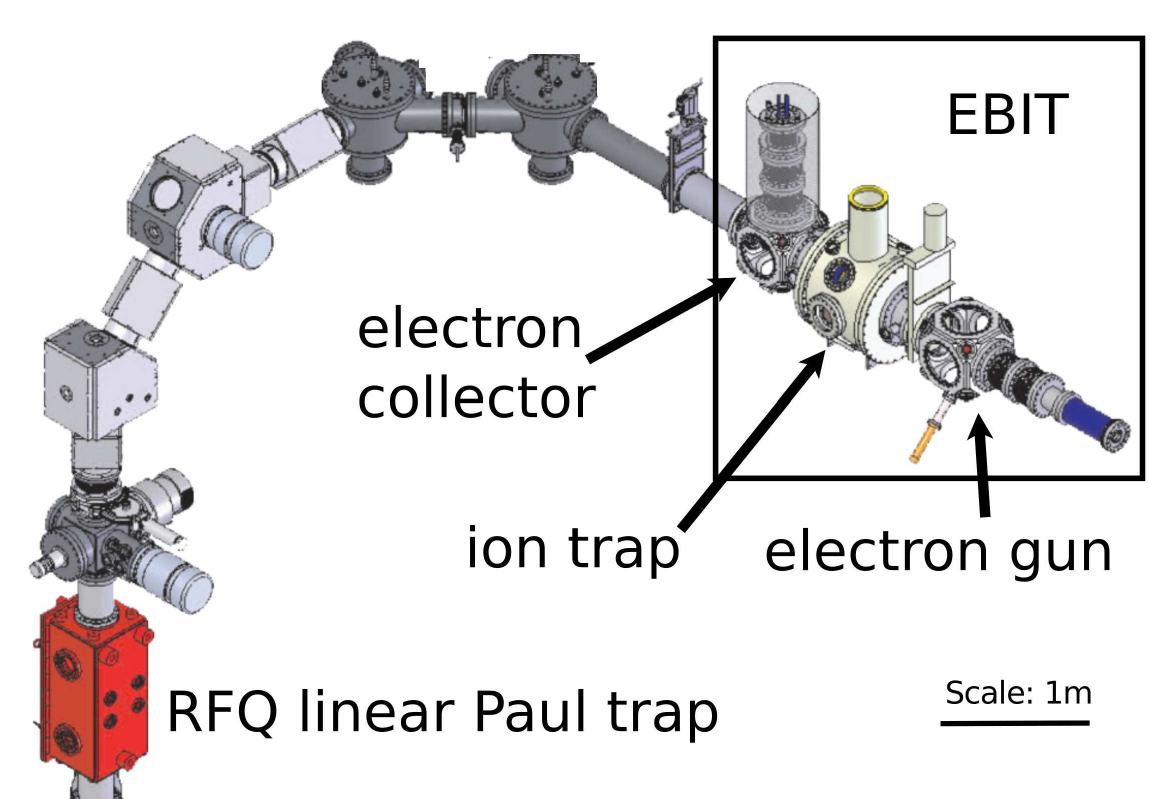}
\caption{A schematic view of the ion traps used in the TITAN decay spectroscopy program.  The ions are extracted in a bunch from the RFQ and injected as singly charged ions to the EBIT where they are stored for charge breeding and decay spectroscopy.  After the ions have been stored in the EBIT, they are extracted and dumped downstream away from the detectors.  The ion-bunch cycling is repeated continuously over the course of an experiment.}
\label{fig-1}
\end{figure}

The TITAN facility is primarily used to perform high-precision Penning-trap mass spectrometry on short-lived radioactive nuclides~\cite{Gal14,Fre13,Gal12,Bro12b}.  TITAN has also recently initiated an extensive experimental program of performing in-trap decay spectroscopy with the EBIT~\cite{Ett09,Bru13,Len14,Lea14}.  A schematic view of the TITAN facility at TRIUMF during decay-spectroscopy experiments is displayed in Fig.~\ref{fig-1}. This article presents a brief description of the decay spectroscopy program at TITAN, as well as recently employed methods of improving the overall photon detection sensitivity.

\section{Decay Spectroscopy Overview}
The TITAN EBIT features seven external access ports which allow for the mounting of photon detectors for performing spectroscopy on trapped ions.  These ports are covered with 0.25~mm thick, $>99\%$ pure, pinhole-free Be windows to provide vacuum isolation, while having a minimal effect on X- and $\gamma$-ray transmission.  For the present physics goals~\cite{Fre07}, each of the seven access ports around the EBIT houses a lithium-drifted silicon (Si(Li)) detector~\cite{Lea14}.  The particular energy region of interest is $E_{\gamma}\leq50$~keV, where X-rays from the electron capture (EC) decay process can be observed.  Since these energies correspond to low-amplitude signals from the Si(Li) detectors, a high level of environmental control is needed to reach the sensitivity required for the observation of weak transitions.
\begin{figure}[t!]
\centering
\includegraphics[width=\columnwidth]{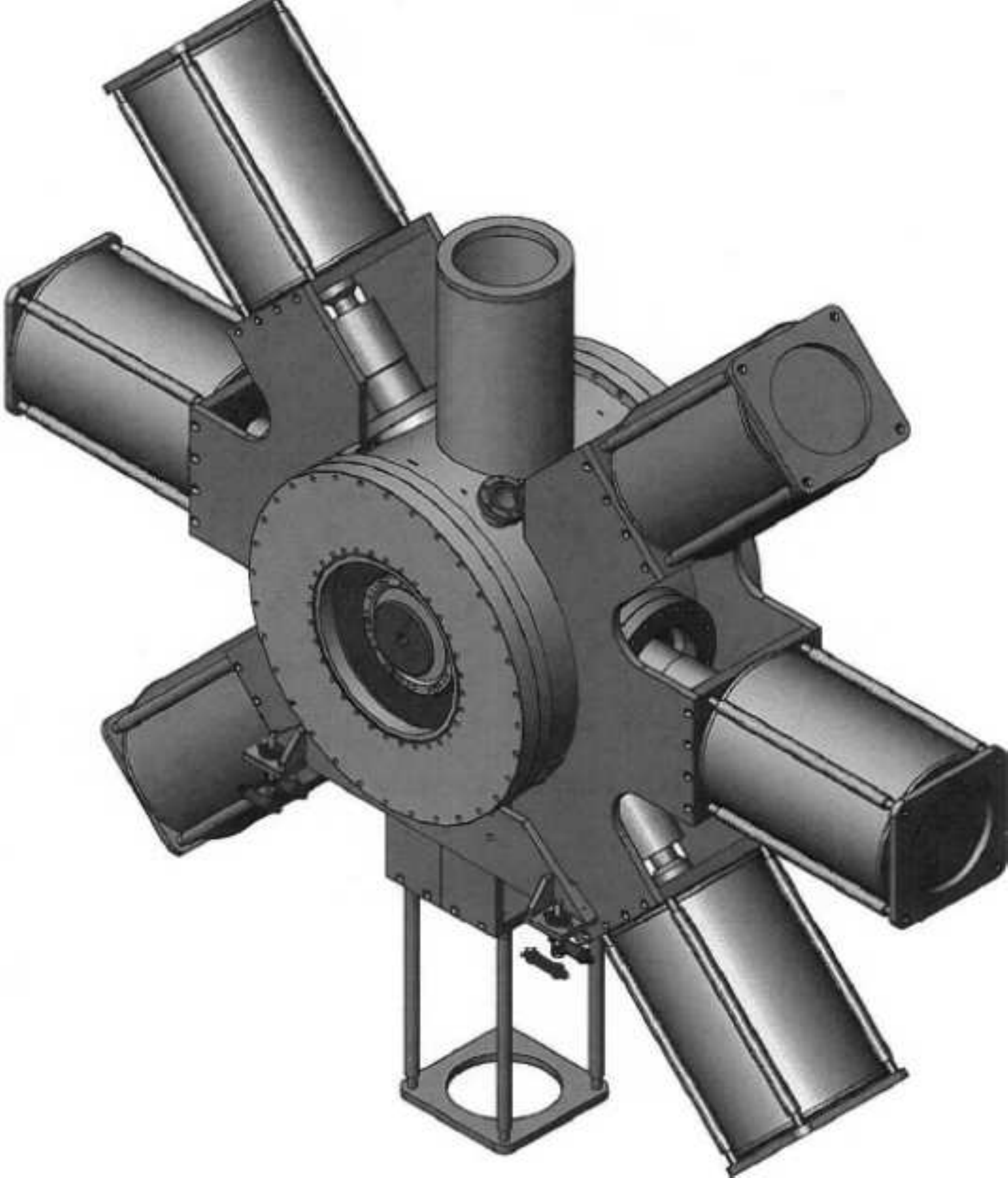}
\caption{The decay-spectroscopy setup surrounding the TITAN EBIT.  Shown are six of the seven detectors (the bottom housing is empty for illustrative purposes) which are supported by an aluminum frame which is attached to the trap.  It is at this connection point where the vibrational noise from the EBIT is transferred to the Si(Li) detectors.}
\label{fig-2}
\end{figure}

\section{Vibrational-Noise Reduction}
The EBIT employs a helium cryo-cooler for the superconducting magnet which supplies high-pressure He gas to the cold-head~\cite{Lap10}.  In this process, the compression cylinder generates vibrational noise.  The photon detectors are located in a frame that is directly mounted to the base of the magnet housing, which transfers the vibrational noise generated by the He compressor directly to the Si(Li) detectors (Fig.~\ref{fig-2}).  These vibrations resonate at many frequencies in the aluminum detector-support frame and generate acoustic noise up to several-hundred Hz.  The distribution of high-frequency vibrations that exist at one of the horizontal access ports is presented in Ref.~\cite{Lea14}.  These vibrations have a significant effect on the signal quality from the Si(Li) detectors, primarily on the observed energy resolution.  The following section presents tests of different damping materials for possible vibration-reduction solutions.
\begin{figure}[t!]
\centering
\includegraphics[width=0.9\linewidth]{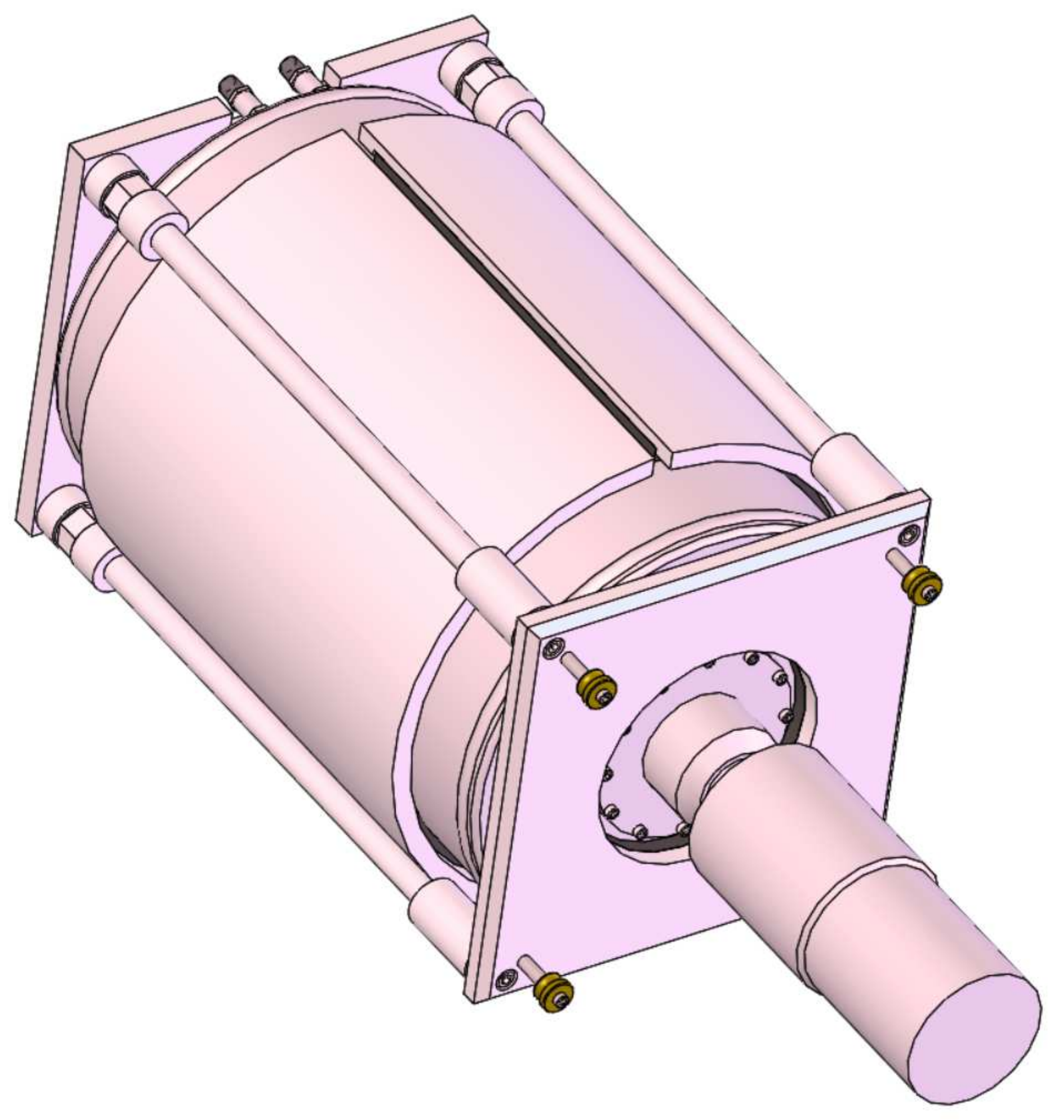}
\caption{A design drawing of a Si(Li) detector in its mounting structure.  Vibration isolation material is located at each contact point, including the front- and back-plates, as well as a ``clam-shell" structure on the sides.  This new housing design for the Si(Li) detectors also improves the versatility of the array, allowing for easier interchangeability between photon detector types.}
\label{fig-3}
\end{figure}

Two primary materials were tested for vibration damping and isolation purposes; {\it Polytech® High Density Polyurethane foam-grade 40} (PHU-40)~\cite{PHU}, and {\it POLYFORM® High Density Molded - Non-Skinned} (PHDM-NS) Polyurethane Foam~\cite{PHDM}.  The PHU-40 material is is a low modulus, open cell Polyurethane foam which exhibits a low compression set, excellent impact resistance, and low permeability when compressed~\cite{PHU}.  The PHDM-NS material is a flexible foam that is characterized by its high resilience, good tear strength, and good elongation properties~\cite{PHDM}.  These materials were selected due to their excellent properties for vibration isolation while under long-term compression.

In order to properly assess the effect of the two isolation materials, detector performances were tested with a $^{133}$Ba source, in four different environments:
\begin{enumerate}
\item {\bf Off-Trap}:  The detector was removed from the trap and placed on a test bench in the ISAC experimental hall.  The results of tests in this environment are considered as the nominal performance.
\item {\bf No Damping}:  The detector was mounted in its housing on the trap with no material between the detector and its mounting points.  This configuration is how all previous measurements were performed.
\item {\bf PHU-40}:  The detector was mounted in its housing on the trap with the PHU-40 material~\cite{PHU} between the detector and its mounting points, as shown in Fig.~\ref{fig-3}.
\item {\bf PHDM-NS}:  The detector was mounted in its housing on the trap with the PHDM-NS material~\cite{PHDM} between the detector and its mounting points, as shown in Fig.~\ref{fig-3}.
\end{enumerate}

\begin{figure}[floatfix!]
\centering
\includegraphics[width=1.15\linewidth]{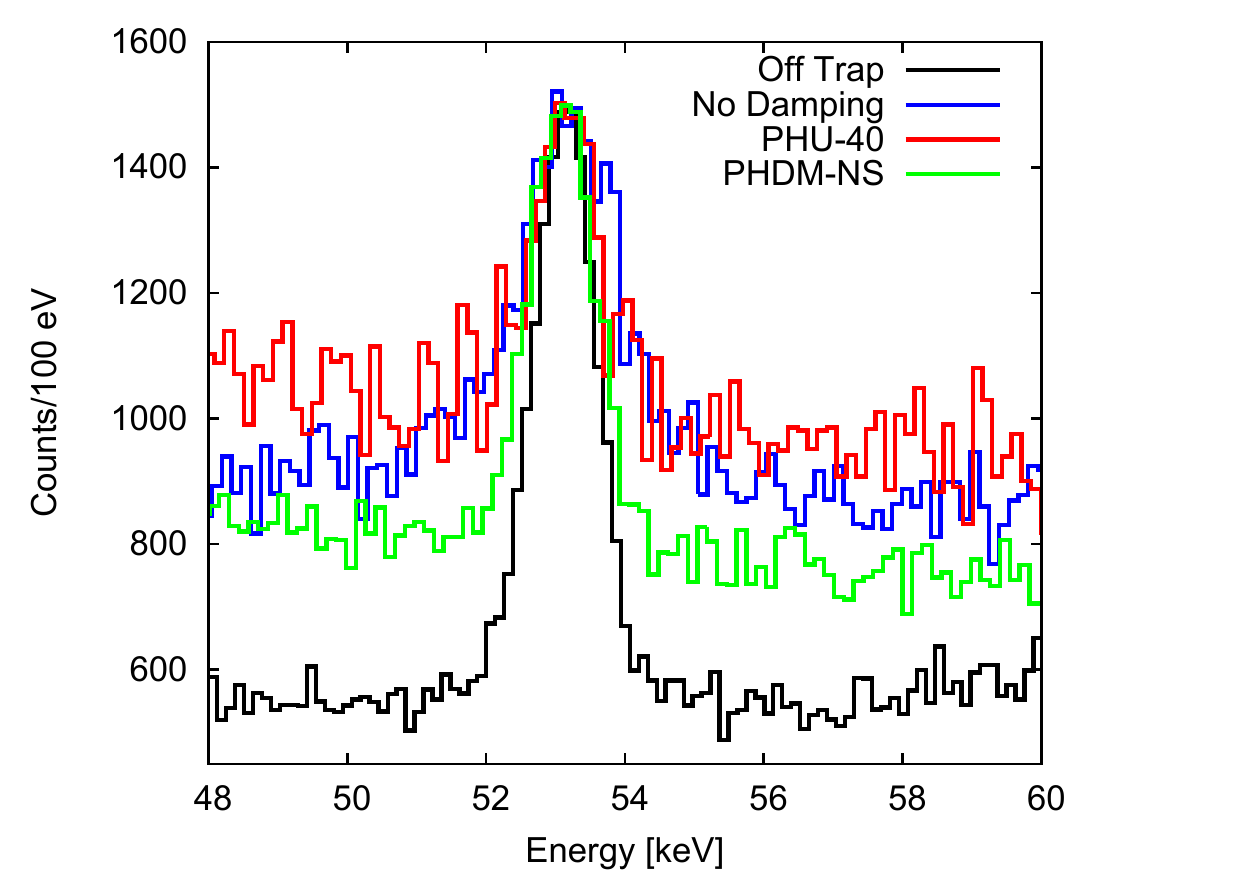}
\caption{A comparison of the observed $^{133}$Ba decay spectra for Detector-427 renormalized by peak height to highlight the resolution difference at 53~keV.  The large peak-to-background difference for the ``Off-Trap" data results from a larger solid-angle coverage of the test bench setup relative to the ``On-Trap" tests.  The list of resulting resolutions for all five detectors is displayed in Table~\ref{table:results}.}
\label{fig-4}
\end{figure}

For each test, the Si(Li) detectors were mounted in a horizontal port on the EBIT, and a collimated $^{133}$Ba source was placed in the port on the opposite side of the trap.  Although this significantly extended the measurement periods required due to the small solid-angle coverage, it was necessary since there is no way of accessing the centre of the trap for calibration purposes (c.f. Ref.~\cite{Lea14}).  Both materials were tested on five of the seven detectors and compared to the detector performance on a test bench.  A comparison of the resulting $^{133}$Ba decay spectra is displayed in Fig.~\ref{fig-4} for one of the five detectors, and the quantitative comparison for all five detectors is shown in Table~\ref{table:results}. 

\section{Active Compton Suppression}
\begin{figure}[t!]
\begin{center}
\includegraphics[width=\linewidth]{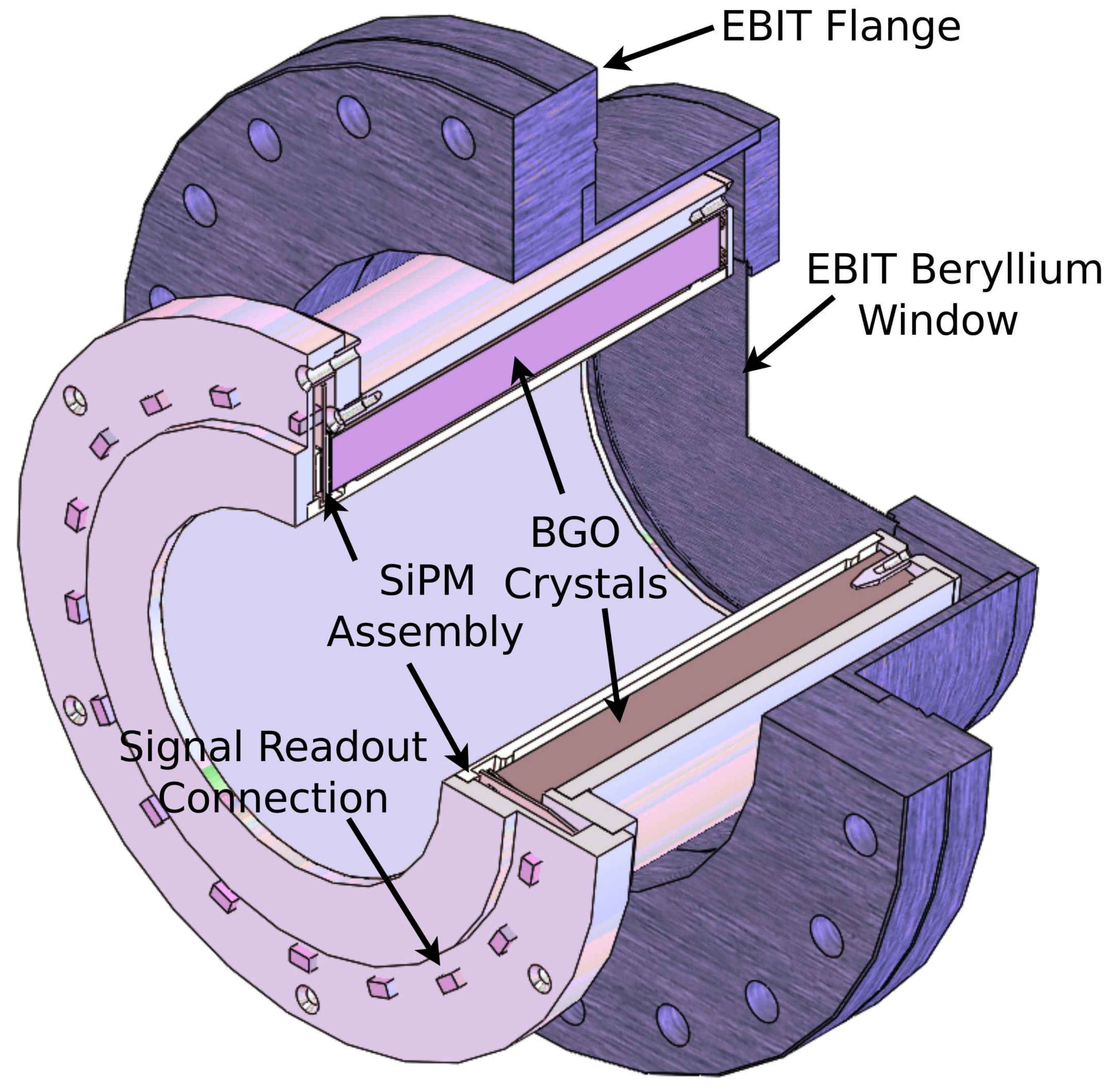}
\end{center}
\caption{A concept design of an active Compton suppression shield solution for the Si(Li) detectors.  The shield itself sits inside the flange on the outside of the EBIT access ports, and each consists of 18 long BGO crystals which surround the detector radially.  The SiPM assembly also includes a Peltier cooling element to keep the thin matrix at temperatures just above $0^\circ$C to reduce dark noise effects.  The detector assembly shown in Fig.~\ref{fig-3} inserts directly into the suppression shield such that the Si(Li) crystal is a few mm away from the thin Be window on the exterior of the EBIT.  This design concept could also be employed for other types of photon detectors that could be coupled to the system in the near future ~\cite{Lea14}.}
\label{fig-5}
\end{figure}

\begin{figure}[t!]
\begin{center}
\includegraphics[width=0.95\linewidth]{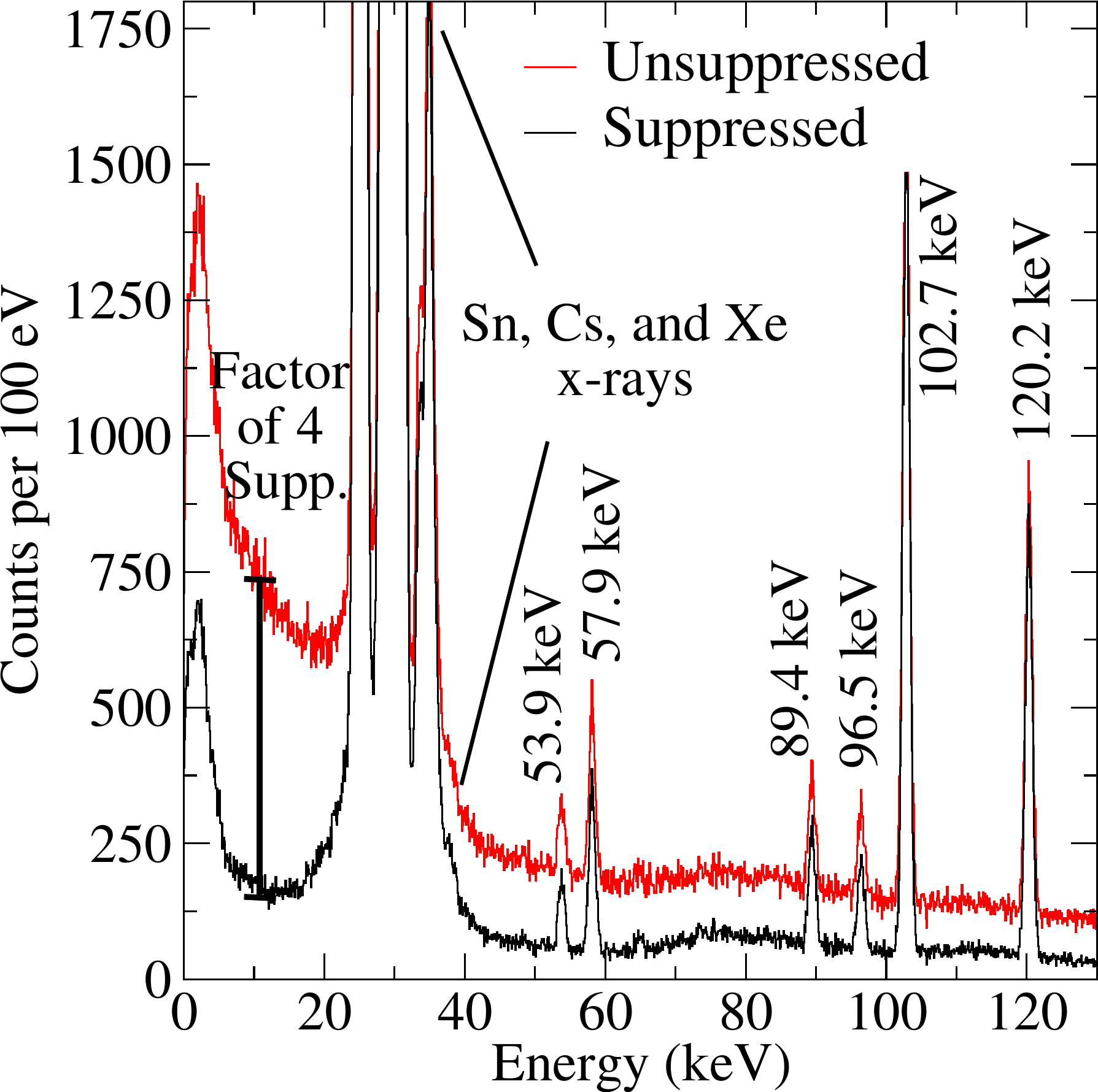}
\end{center}
\caption{A {\sc geant4}~\cite{GEANT} simulation showing the effect of active Compton suppression with BGO shields, as described in the text.  The simulated curves results from the same number of events from one detector during the commissioning $^{124}$Cs experiment~\cite{Len14}, where the red curve is the result for passive Cu and Pb shields only, and the black line is Compton suppressed photon spectrum using 1~cm of BGO surrounding the detector.  The black bar demonstrates the factor-of-four Compton background suppression at 10~keV.}
\label{fig-6}
\end{figure}

\begin{table}[t!]
\centering
\caption{The effect of both damping materials on the observed resolution of the 53~keV $\gamma$-ray following the decay of $^{133}$Ba for five Si(Li) detectors, as described in the text.  The resolution values are given as full-width at half-maximum (FWHM).  In nearly all cases, the nominal resolution vales are achieved using the vibration isolation material PHDM-NS, as highlighted in bold.} 
\label{table:results} 
\begin{tabular}{c|cccc} 
\hline
Det. & Off-Trap & No Damp. & PHU-40 & PHDM-NS \\
No. & (\%) & (\%) & (\%) & (\%)\\
\hline 
404 & \textbf{1.6(1)} & 1.9(1) & 1.7(1) & \textbf{1.6(1)} \\
427 & \textbf{2.0(1)} & 3.5(1) & 2.4(1) & \textbf{2.1(1)} \\
431 & \textbf{2.0(1)} & 2.8(1) & 2.7(1) & \textbf{2.4(1)} \\
432 & \textbf{2.1(1)} & 2.7(1) & 2.2(1) & \textbf{2.0(1)} \\
438 & \textbf{1.9(1)} & 2.4(1) & 2.4(1) & \textbf{2.2(1)} \\
\hline
\end{tabular}
\end{table}
An increase in detection sensitivity is also possible through the elimination of Compton scattered events in the Si(Li) detectors.  This background is generated when the emitted photon interacts with the detector, but does not deposit all of its energy in the crystal, and exits the volume.  The most common method for active Compton background suppression uses a scintillating material with a high nuclear charge ($Z$) that surrounds the photon detector.  This method improves the effective signal-to-background ratio by vetoing events with incomplete energy deposition in the Si(Li) crystal.  Typically, the inorganic scintillating crystal Bismuth Germanium Oxide (BGO) is used in combination with photo-multiplier tubes (PMTs) to achieve this suppression~\cite{KNOLL}.  The TITAN EBIT, however, poses a considerable problem since PMTs are unable to operate in the up to $\sim0.15$~T fringe magnetic fields.  To circumvent this issue, development has begun to couple the BGO crystals to silicon photo-multipliers (SiPMs) instead of PMTs, which are able to operate in a high-field environment.  In the current design (Fig.~\ref{fig-5}), the SiPMs are individually coupled to eighteen 1~cm thick bars of BGO which surround the Si(Li) detector.  This ensures optical isolation of the individual components, as well as decreasing the overall size of the crystals needed.  Using this design, a factor of 2-4 suppression of Compton-generated background at low energies has been demonstrated in Monte-Carlo-based simulations (Fig.~\ref{fig-6}).  Further work on the electronic readout of the SiPMs is required before it can be effectively coupled to the BGO crystals (due to their low light output), and is presently underway~\cite{RetPC}.

\section{Conclusions}
In summary, an in-trap decay spectroscopy tool has been developed at TRIUMF using TITAN's electron-beam ion-trap.  The ion-trap environment allows for the detection of low-energy photons by providing backing-free storage, while simultaneously guiding charged decay particles away from the trap centre via the strong magnetic field.  To further increase the scientific reach of this facility, two studies were performed to improve the overall sensitivity for photon detection.

A study to examine the effect of vibration isolation of the Si(Li) detectors was needed to removed the induced mechanical noise which results from contact with the EBIT cryo-cooler pump.  The best performance was achieved using the POLYFORM® PHDM-NS material.  A study on the implementation of Compton-scattering suppression shields was also performed.  The concept of using BGO crystals coupled to SiPM readouts was used, since the high-field environment of the EBIT prevents the use of standard PMTs.  Preliminary estimates from {\sc geant4} simulations suggest that the Compton suppression could reduce the observed background by up-to a factor of 4.  The implementation of both methods of background reduction is currently underway, and will provide a significant improvement for both photon detection sensitivity and energy resolution.

\section{Acknowledgements}
TRIUMF receives federal funding via a contribution agreement with the National Research Council of Canada (NRC).  This work was partially supported by the Natural Sciences and Engineering Research Council of Canada (NSERC), and the Deutsche Forschungsgemeinschaft (DFG) under grant FR 601/3-1.  The authors would also like to thank Fabrice Reti\'ere and the TRIUMF Science Technology Department for their collaborative effort in development of the SiPM detectors.  KGL thanks Stephan Ettenauer and Thomas Brunner for useful discussions regarding this work.

%
%
%

\end{document}